\documentclass[aps,prl,reprint,superscriptaddress,longbibliography,10pt]{revtex4-2}
\usepackage{amsmath, amssymb, graphicx, xcolor, braket,mathtools}
\usepackage{bbold, bm}
\usepackage{verbatim}
\usepackage{bbm}
\usepackage[most]{tcolorbox}

\begin{document}

\setlength{\dbltextfloatsep}{8pt plus 0.1pt minus 2pt}

\title{Symmetry-Induced Weyl Nodes in Interacting Multi-Terminal Josephson Junctions}

\author{Peter Zalom}
\email{zalomp@fzu.cz}
\affiliation{Institute of Physics, Czech Academy of Sciences, Na Slovance 2, CZ-18200 Praha 8, Czech Republic}

\author{Gustavo Diniz}
\affiliation{Institute of Physics, Czech Academy of Sciences, Na Slovance 2, CZ-18200 Praha 8, Czech Republic}

\author{David Christian Ohnmacht}
\altaffiliation[Present address: ]{
    QuTech and Kavli Institute of Nanoscience,
    Delft University of Technology,
    2628 CJ Delft, The Netherlands
}
\affiliation{Fachbereich Physik, Universit\"at Konstanz, D-78457 Konstanz, Germany}

\author{Wolfgang Belzig}
\email{wolfgang.belzig@uni-konstanz.de}
\affiliation{Fachbereich Physik, Universit\"at Konstanz, D-78457 Konstanz, Germany}

\date{\today}

\begin{abstract}
We show that an emergent geometric symmetry generates non-trivial topology in quantum-dot-based multiterminal Josephson junctions. It confines same-spin Andreev bound state crossings to an analytic one-dimensional manifold of the synthetic Brillouin zone, where interdot coupling selects Weyl nodes of charge $\pm1$ in the singlet sector and doubly degenerate cones of charge $\pm2$ in the doublet sector, at gate-tunable locations. The mechanism yields a spectroscopic detection protocol and a design principle for fabricating devices with non-trivial topological signatures.
\end{abstract}


\maketitle

\noindent\textit{Introduction.}---Non-trivial band topology can emerge from non-generic Hamiltonian properties alone~\cite{Nielsen-1983}, yet in many quantum systems it is generated by an underlying symmetry~\cite{Kane-2005,Fu-2007}. Though rooted in crystalline band theory, topology admits a far more general formulation, where any compact parameter space serves as a synthetic Brillouin zone (sBZ), with Berry curvature~\cite{Berry-1984} and Chern numbers~\cite{Thouless-1982} defined through parametric evolution of quantum states. The in-gap Andreev bound states (ABS) of multiterminal Josephson junctions (MTJJs) constitute such synthetic bands~\cite{Riwar-2016,Zazunov-2017,Repin-2022}, which are predicted to be topologically active in a number of settings~\cite{Eriksson-2017,Klees-2020,Teshler-2023,Meyer-2017b}. While scattering-matrix approaches have clarified some constraints in non-interacting MTJJs~\cite{Schnyder-2008,Xie-2018}, systematic predictions of when and why MTJJ system exhibit topologically non-trivial features are missing.

Moreover, unambiguous topological signatures in MTJJs have remained elusive despite experimental confirmation of other hallmark phenomena, including coherent multi-Cooper pair transport~\cite{Freyn-2011,Cohen-2018,Arnault-2022,Ohnmacht-2024f,Arnault-2025}, the  multi-terminal Josephson effect~\cite{Pankratova-2020,Kahn-2025}, the Josephson diode effect~\cite{Coraiola-2024}, and ABS band engineering~\cite{Coraiola-2023,Coraiola2024}. To facilitate the detection and practical use of topological singularities, current research focuses on Josephson junctions with quantum dots (QDs) due to their high degree of gate-tunability. Indeed, recent four-terminal spectroscopy accessed the synthetic 3D Andreev spectrum of a triple-dot Andreev molecule, where the underlying effective model predicted non-trivial topology~\cite{Antonelli2025h}. Compounding this, in interacting MTJJs Weyl nodes appear principally as excited crossings in analogy to the doublet chimney effect of ordinary QD-based junctions~\cite{Pavesic-2024,Zalom-2024prl,Rolih-2026}. 

In this Letter we overcome these obstacles by identifying geometric symmetries $\mathcal{P}_{\alpha}$ intrinsic to multi-dot MTJJs. These define analytical $\bm{\chi}$-manifolds in the sBZ, to which topological singularities are pinned at gate- and flux-tunable locations. These findings enable targeted  protocols for detecting and systematic control of topological singularities in multi-dot MTJJs.

\begin{figure}[t]
	\includegraphics[width=0.9\columnwidth]{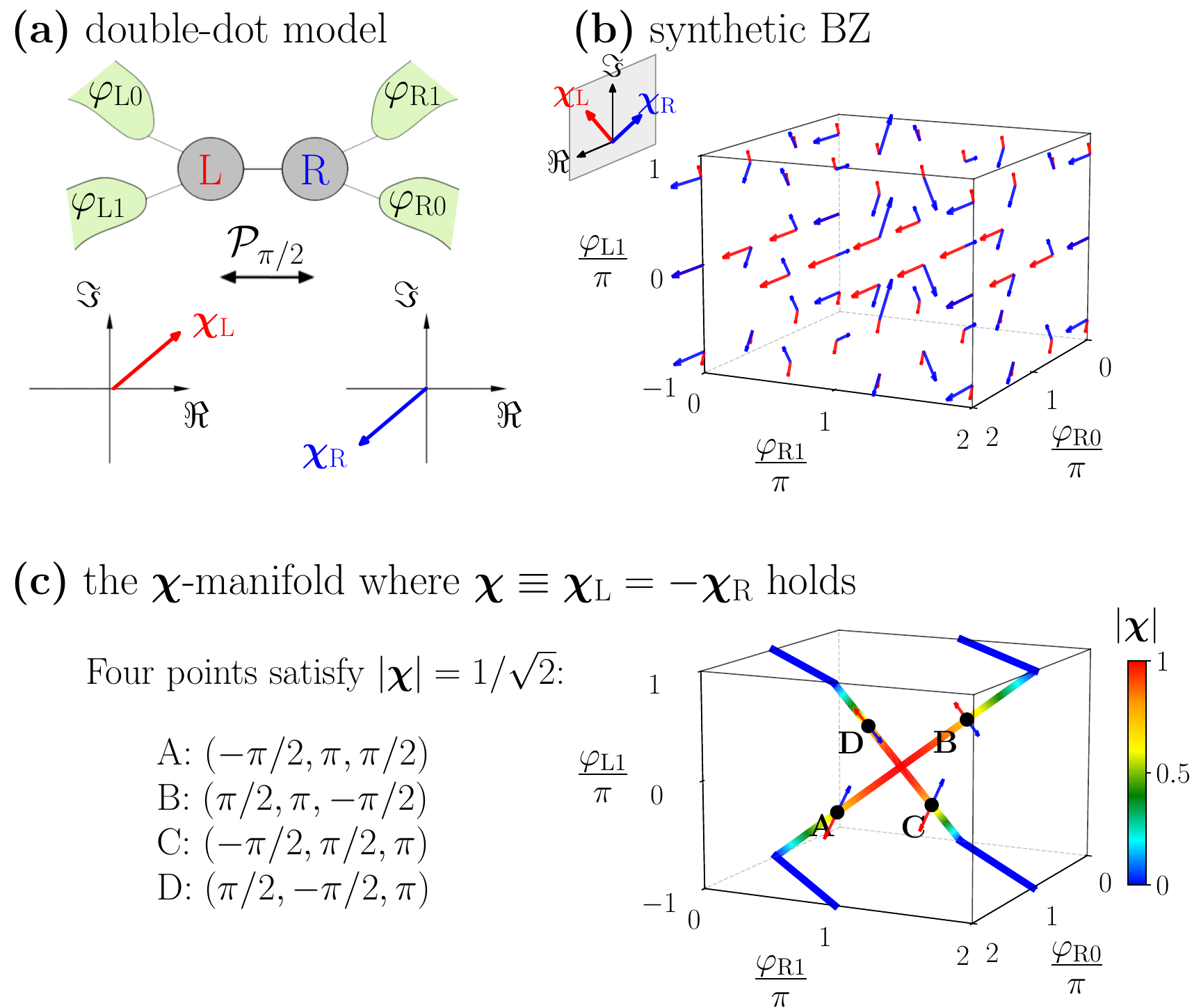}
	\vspace{-3mm}
	\caption{
		Symmetry $\mathcal{P}_{\pi/2}$ in double-dot model~\eqref{eq:Hdqd}:
        (a) 
		SC environments of left ($i=\mathrm{L}$) and right ($i=\mathrm{R}$) dots are fully defined by total coupling strengths $\Gamma_i\equiv\Gamma_{i0}+\Gamma_{i1}$ and geometric factors $\bm{\chi}_i$. 		
		(b)
		In the gauge $\varphi_{\mathrm{L}0}=0$, phases $\varphi_{\mathrm{L}1}$, $\varphi_{\mathrm{R}0}$, $\varphi_{\mathrm{R}1}$ span the sBZ where pairs of $(\bm{\chi}_{\mathrm L},\bm{\chi}_{\mathrm R})$ are assigned.
		(c)
		$\mathcal{P}_{\pi/2}$ can only be satisfied on the $\bm{\chi}$-manifold defined by $\bm{\chi}_{\mathrm L}=-\bm{\chi}_{\mathrm R}$. The example has all dot-lead couplings the same. The bar legend shows $|\bm{\chi}|=|\bm{\chi}_{\mathrm L}|=|\bm{\chi}_{\mathrm R}|$ on the manifold. Blue-colored lines with $|\bm{\chi}|=0$ form nodal lines in the sBZ.  
		\label{fig:combined}
	}
\end{figure}

\textit{Models.}---We consider QD-based MTJJs within experimentally realistic constraints thereby excluding direct inter-terminal couplings as proposed in~\cite{Klees-2020}. Since single-dot MTJJs without such couplings map onto topologically trivial counterparts~\cite{Zalom-2024prl}, multi-dot systems are then required. The simplest such system is the double-dot JJ with four superconducting (SC) terminals~\cite{Teshler-2023}, where $i=\mathrm{L}$ labels the left and $i=\mathrm{R}$ the right QD while $l \in \{0,1\}$ tracks the leads attached to each QD as shown in Fig.~\ref{fig:combined}(a). The system obeys the Hamiltonian
\begin{align}
H_2
&=
H_{\mathrm{d}}
+
\sum_{il}
H_{\mathrm{T}, il}
+
\sum_{il}
H_{\mathrm{SC}, il},
\label{eq:Hdqd}
\\
H_{\mathrm{d}}
&=
\sum_{\sigma i} 
\left(
\delta_{i} - \frac{U_i}{2}
\right)
d^{\dagger}_{\sigma i}
d^{\vphantom{\dagger}}_{\sigma i}
+
U_i
d^{\dagger}_{\uparrow i}
d^{\vphantom{\dagger}}_{\uparrow i}
d^{\dagger}_{\downarrow i}
d^{\vphantom{\dagger}}_{\downarrow i}
\nonumber
\\
&+
\sum_{\sigma} 
t_{\mathrm{LR}}
\left(
d^{\dagger}_{\sigma \mathrm{L}}
d^{\vphantom{\dagger}}_{\sigma \mathrm{R}}
+
d^{\dagger}_{\sigma \mathrm{R}}
d^{\vphantom{\dagger}}_{\sigma \mathrm{L}}
\right),
\label{eq:Hdot}
\end{align}
where $d^{\dagger}_{\sigma i}$ creates a fermion of spin projection $\sigma$ on dot $i$, $U_i$ is the on-site Coulomb repulsion, $\varepsilon_{\mathrm{d},i}$ is the energy level of dot $i$ and $\delta_i \equiv \varepsilon_{\mathrm{d},i} + U_i/2$. Interdot coupling $t_{\mathrm{LR}}$ is set real by gauge freedom. Each dot couples to two SC leads via tunneling Hamiltonians $H_{\mathrm{T}, il}$ (for details see the supplemental material, short: SM \cite{[{See Supplemental Material at }][{ for details on model definitions, Green's function derivation, Al theory, calculation of velocity matrices and additional data for coupling asymmetric scenario which includes Refs.~\cite{Zalom-2024prl,Zalom-2021, Zonda-2023, Klees-2020,Teshler-2023,Liu-2016,dmrg,itensor,Satori-1992,Ohnmacht-2025r}.}]supp}) with energy-independent dot-lead coupling strengths $\Gamma_{il}$. All SC terminals obey Bardeen-Cooper-Schrieffer Hamiltonians $H_{\mathrm{SC}, il}$ with the same gap $|\Delta_{il}|=\Delta$, reflecting the common experimental setting (see SM). Using the gauge $\varphi_{\mathrm{L}0} \equiv 0$ throughout this Letter, the sBZ is directly spanned by the phases $\varphi_{\mathrm{L} 1}$, $\varphi_{\mathrm{R} 0}$ and $\varphi_{\mathrm{R} 1}$ [Fig.~\ref{fig:combined}(b)]. The generalization to three or more dots is defined in the SM.

\noindent\textit{Emergent geometric symmetries.}---The many-body spectrum of model~\eqref{eq:Hdqd} is fully encoded in the determinant of the inverse dot-sector Green's function. Any operation leaving it invariant is thus a spectral symmetry, even if not manifest directly in the Hamiltonian~\cite{Zalom-2024prl}. To identify such operations, we introduce the Nambu spinor ${D^\dagger = (d^\dagger_{\mathrm{L},\uparrow}, d_{\mathrm{L},\downarrow}, d^\dagger_{\mathrm{R},\uparrow}, d_{\mathrm{R},\downarrow})}$ and solve for the non-interacting ($U_{\mathrm{L}} \!=\! U_{\mathrm{R}} \!=\! 0$) retarded Green's function of the dot sector (details in SM):
\begin{align}
\mathbb{G}_{0,\mathrm{d}}^{-1}(\omega^+)
=
\omega^+ \!\mathbb{1} 
-
\begin{bmatrix} 	
	 \mathbb{E}_{\mathrm L} \!+\!\mathbb{\Sigma}_{\mathrm L}(\omega^+)  & \mathbb{T}_{\mathrm{LR}}
	\\
	\mathbb{T}_{\mathrm{LR}}  &  \mathbb{E}_\mathrm{R}  \!+\!\mathbb{\Sigma}_\mathrm{R}(\omega^+)
\end{bmatrix}.
\label{eq:G0}
\end{align}
Here, $\mathbb{T}_{\mathrm{LR}}=t_{\mathrm{LR}}\sigma_z$, $\mathbb{E}_i=(\delta_i-U_i/2)\sigma_z$ with $\sigma_{x,y,z}$ the Pauli matrices, and since no terminal is shared among the dots, the SC environments contribute only block-diagonal self-energies $\mathbb{\Sigma}_i(\omega^+)$ which in accord with~\cite{Zalom-2021} read
\begin{align}
	\mathbb{\Sigma}_i (\omega^+)
	=
	\Gamma_i 
	\begin{pmatrix} 	
		\omega^+ & \bm{\chi}_i
		\\ 
		\bm{\chi}^*_i	& \omega^+
	\end{pmatrix}
	F(\omega^+),
	\label{eq:SigmaLR}
\end{align}
where $F(\omega^+,\Delta)$ is a universal function detailed in the SM. The SC environment of dot $i$ is then uniquely characterized via a pair of invariants: the total tunneling strengths $\Gamma_i\equiv\Gamma_{i0}+\Gamma_{i1}$ and geometric factors
\begin{align}
	\bm{\chi}_i
	&\equiv
	\left(
	\Gamma_{i0} 
	e^{i \varphi_{i0}} 
	+ 
	\Gamma_{i1} 
	e^{i \varphi_{i1}}
	\right)
	/\Gamma_{i},
	\label{eq:geoms}
\end{align}
that define complex-valued fields in the sBZ as shown in Fig.~\ref{fig:combined}(b). When $\delta_{\mathrm L}=\delta_{\mathrm R}$ and $\Gamma_{\mathrm L}=\Gamma_{\mathrm{R}}$, the non-interacting Green's function~\eqref{eq:G0} is invariant under 
\begin{align} 
		\mathcal{P}_{\alpha}^{\dagger}
		d^{\dagger}_{\sigma \mathrm{L}} 
		\mathcal{P}_{\alpha}
		=
		e^{i\alpha}	
		d^{\dagger}_{\sigma \mathrm{R}},
		\qquad
		\mathcal{P}_{\alpha}^{\dagger}
		d^{\dagger}_{\sigma \mathrm{R}} 
		\mathcal{P}_{\alpha}
		=
		e^{i\alpha}	
		d^{\dagger}_{\sigma \mathrm{L}},
		\label{eq:unitary}
\end{align}
for $\bm{\chi}_{\mathrm L} \!=\! \bm{\chi}_{\mathrm R}$ with $\alpha=0,\pi$ and for $\bm{\chi} \! \equiv \! \bm{\chi}_{\mathrm L} \!=\! -\bm{\chi}_{\mathrm R}$ with $\alpha \!=\! \pm\pi/2$. As seen directly from the Coulomb terms of~\eqref{eq:Hdqd}, both symmetries extend to the fully interacting scenario once $U_{\mathrm{L}}=U_{\mathrm{R}}$. However, for $\alpha=0,\pi$ the lowest states of the decoupled subspaces cross only at $t_{\mathrm{LR}} = 0$ (see SM). Consequently, only $\mathcal{P}_{\pi/2}$ induces non-trivial crossings and we henceforth concentrate only on it. 
 
\begin{figure*}[t]
	\includegraphics[width=2.0\columnwidth]{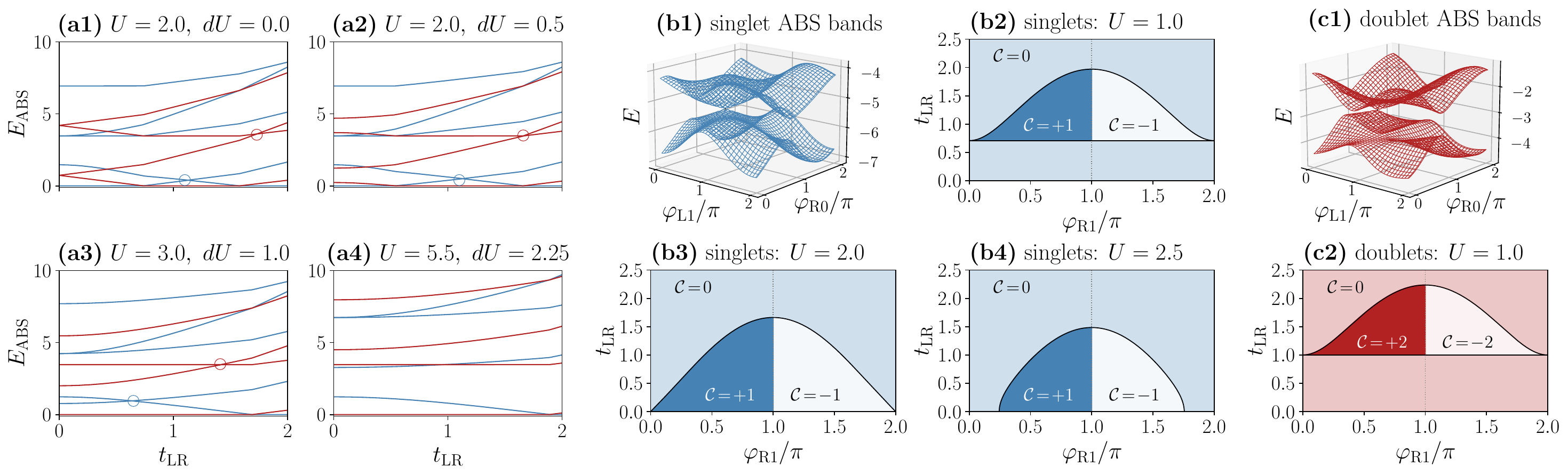}
	\vspace{-4mm}
\caption{
	AL theory results:
    (a$1$)--(a$4$) 
	Sweeping $t_{\mathrm{LR}}$ on the $\bm{\chi}$-manifold induces isolated singlet (blue circles) and doublet (red circles) band crossings via~\eqref{eq:sing_cross} and~\eqref{eq:doub_cross}. Singlet crossings survive finite $dU$ but disappear beyond a $U$-threshold~(a4); doublet crossings are $U$-independent.
	(b$1$), (c$1$) 
	Weyl cone and doubly degenerate Weyl cone formed by the two lowest singlet and doublet bands respectively in the $\varphi_{\mathrm{L} 1}$-$\varphi_{\mathrm{R} 0}$ plane at $\varphi_{\mathrm{R} 1}=-\pi/2$. 
	(b$2$)--(b$4$), (c$2$) 
	Analytic topological phase diagrams from~\eqref{eq:sing_cross} and~\eqref{eq:doub_cross} for singlet and doublet ABS bands respectively (details in End Matter) showing topological phases of finite Chern numbers $\mathcal{C}=\pm1$ and $\mathcal{C}=\pm2$ respectively.
	All panels: $\Gamma_{il}=\delta=1$, in (a$1$)--(a$4$): $|\bm{\chi}|=1/\sqrt{2}$; in (b$1$), (c$1$): $U_{\mathrm{L}}=U_{\mathrm{R}}=2$.
	\label{fig:al_results}
}
\end{figure*}

Crucially, Eq.~\eqref{eq:unitary} targets geometric factors via ${\bm{\chi} \! \equiv \! \bm{\chi}_{\mathrm L} \!=\! -\bm{\chi}_{\mathrm R}}$ which defines a submanifold in the sBZ where $\mathcal{P}_{\pi/2}$ is exclusively satisfied by setting $\delta \!\equiv\! \delta_{\mathrm L} \!=\!\delta_{\mathrm R}$, $\Gamma\!=\!\Gamma_{\mathrm L}\!=\!\Gamma_{\mathrm{R}}$ and $U\!\equiv\! U_{\mathrm{L}} \!=\! U_{\mathrm{R}}$. On such a $\bm{\chi}$-manifold [Fig.~\ref{fig:combined}(c)], each spin multiplet then decomposes into orthogonal subspaces of parity $P=\pm$, allowing same-spin states of opposite parity to cross. Whether they actually do is, however, governed by the interdot coupling $t_{\mathrm{LR}}$, which can be swept without breaking $\mathcal{P}_{\pi/2}$. 

A given $t_{\mathrm{LR}}$ generically selects crossings at discrete values of $|\chi|$ which thus become isolated on the $\bm{\chi}$-manifold [see for example the points $A$-$D$ in Fig.~\ref{fig:combined}(c)] which guarantees generically linear gap opening along the manifold itself. The only exception are $|\bm{\chi}|=0$ lines, where a given $t_{\mathrm{LR}}$ induces crossings for all points or not at all. Consequently, these control the emergence of nodal lines in the sBZ due to $\mathcal{P}_{\pi/2}$. Moving away from the $\bm{\chi}$-manifold breaks $\mathcal{P}_{\pi/2}$. The ensuing mixing of the parity sectors then generically gaps out the crossings linearly~\footnote{The gap opening is generically linear since mutual cancellation of all linear terms even in one particular direction of sBZ would require explicit constraints in~\eqref{eq:Hdqd} which emerge at $\delta_{\mathrm L}=\delta_{\mathrm R}=0$, for example.} for all values of $\bm{\chi}$. Consequently, $\bm{\chi}\neq0$ points generically satisfy all defining features of Weyl nodes on the $\bm{\chi}$-manifold, which thus represents a pinning locus of topological singularities that migrate along it as $t_{\mathrm{LR}}$ is swept.

\textit{Atomic limit.}---On the $\bm{\chi}$-manifold, avoided crossings of same-spin ABS bands are lifted by $\mathcal{P}_{\pi/2}$. The remaining question is which microscopic mechanism drives the crossings and under which conditions linear gap opening is guaranteed. To expose this analytically, we first turn to the atomic limit (AL), which imposes $\Delta\rightarrow\infty$ at a fixed ratio of $\Delta$ to the SC band width~\cite{Meng-2009}. The tunneling self-energies~\eqref{eq:SigmaLR} then become $\omega$-independent rendering the problem equivalent to a two-site Hamiltonian
\begin{align}
H_{2,\mathrm{AL}}
=
H_{\mathrm{d}}
+
\sum_{i}
\Gamma_i  
\left(
\bm{\chi}_i
d^{\dagger}_{\uparrow i}
d^{\dagger}_{\downarrow i}
+
\bm{\chi}_i^*
d^{\vphantom{\dagger}}_{\downarrow i}
d^{\vphantom{\dagger}}_{\uparrow i}
\right).
\label{eq:al_ham_dqd}
\end{align}
Spin-rotation symmetry diagonalizes Hamiltonian~\eqref{eq:al_ham_dqd} into blocks $H^{\mathrm x}_{2,\mathrm{AL}}$ with $x$ denoting spin $S$ and its projection $S_z$. Hamiltonian~\eqref{eq:al_ham_dqd} possesses $5$ singlets ($x=s$), $4$ doublets ($S=1/2$) each with $S_z=1/2$ ($x=\,\uparrow$) and $S_z=-1/2$ ($x=\, \downarrow$) projections and one triplet ($S=1$) that has no counterpart to cross with and remains topologically trivial. For $x=s,\uparrow,\downarrow$, parity $P$ enforces 
\begin{align}
H^{\mathrm x}_{2,\mathrm{AL}}
=
\begin{pmatrix}
	H_+^{\mathrm x} & H_{+-}^{\mathrm x}
	\\
	H_{-+}^{\mathrm x} & H_-^{\mathrm x}
\end{pmatrix}
\label{eq:blocks}
\end{align}
on the $\bm{\chi}$-manifold. The off-diagonal blocks vanish always when $\delta_{\mathrm L}=\delta_{\mathrm R}$ and ${dU\equiv(U_{\mathrm{L}}-U_{\mathrm{R}})/4 = 0}$ as a result of $\mathcal{P}_{\pi/2}$ being satisfied. Nevertheless, \eqref{eq:blocks} facilitates analytic calculation of same-spin ABS band crossings also when interaction asymmetries $dU \neq 0$ break $\mathcal{P}_{\pi/2}$. 

In the singlet sector, off-diagonal blocks of~\eqref{eq:blocks} are independent of $dU$. When $\delta_{\mathrm L}=\delta_{\mathrm R}$ on the $\bm{\chi}$-manifold, the singlet subspace decomposes  into a $3 \times 3$ singlet block $H_+^{\mathrm s}$ and a  $2 \times 2$ block $H_-^{\mathrm s}$. The former yields eigenenergies $2\delta$, $2\delta \pm 2\sqrt{\delta^2 + |\Gamma\bm{\chi}|^2}$ while the latter gives $(2\delta-U/2) \pm \sqrt{4|t_{\mathrm{LR}}|^2+U^2/4}$. The two lowest singlets with opposite parities therefore cross at 
\begin{align}
	|t_{\mathrm{LR},\mathrm{s}}|
	\!=\!
	\sqrt{
		\delta^2 
		+
		|\Gamma \bm{\chi}|^2 
		- 
		(U/2) \sqrt{(\delta^2+|\Gamma \bm{\chi}|^2)}
	}
	\label{eq:sing_cross}
\end{align}
as marked in Figs.~\ref{fig:al_results}(a1)-(a3) by blue circles. Eq.~\eqref{eq:sing_cross} represents a microscopic mechanism of how $t_{\mathrm{LR}}$ selects a unique ${|\bm{\chi}|\!=\!|\bm{\chi}_{\mathrm L}|\!=\!|\bm{\chi}_{\mathrm R}|}$ and thus isolated singlet crossing on the $\bm{\chi}$-manifold. Increasing $U/2 \rightarrow \sqrt{(\delta^2+|\Gamma \bm{\chi}|^2)}$ in \eqref{eq:sing_cross} pushes $|t_{\mathrm{LR},\mathrm{s}}|\rightarrow0$ until singlet crossings vanish [Fig.~\ref{fig:al_results}(a4)]. Notably, singlet crossings~\eqref{eq:sing_cross} are not affected by interaction asymmetries since $H^{\mathrm s}_{2,\mathrm{AL}}$ including its off-diagonal blocks is $dU$-independent. 

\begin{figure*}[t]
	\includegraphics[width=2.07\columnwidth]{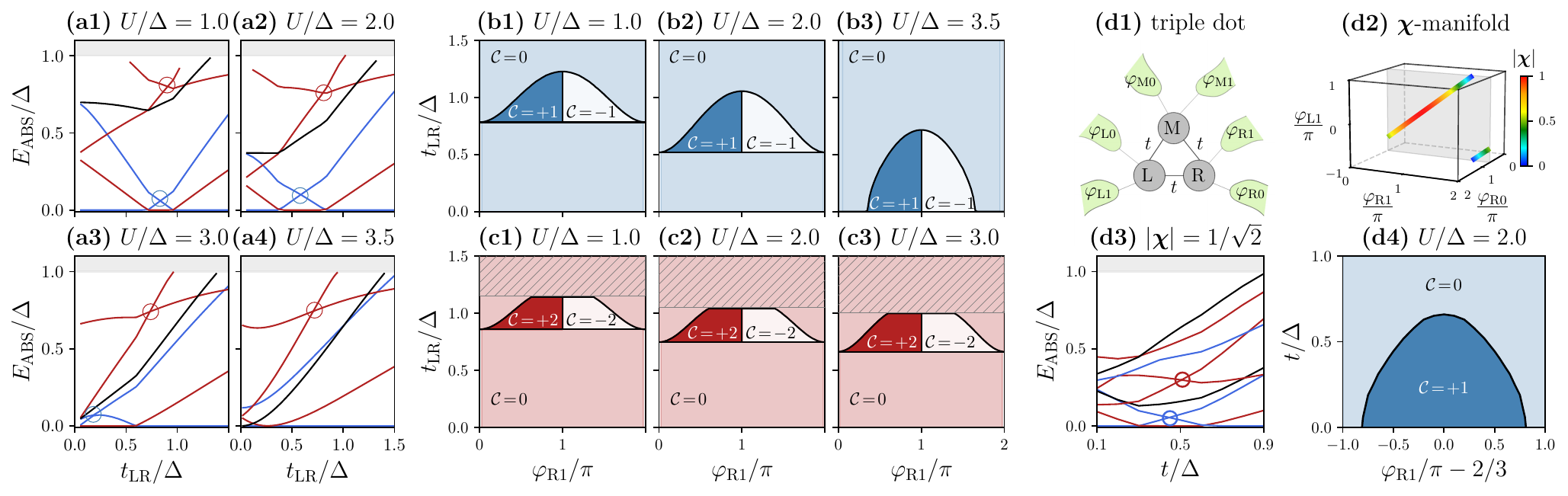}
	\vspace{-7mm}
	\caption{
        DMRG results:
		(a$1$)--(a$4$)
		Sweeping $t_{\mathrm{LR}}$ at $|\bm{\chi}|\approx 0.31$ with $\Gamma_{\mathrm L}\!=\!\Gamma_{\mathrm{R}}\!=\!\Delta$ and $\delta_{\mathrm L}\!=\!\delta_{\mathrm R}=\Delta$ reveals singlet (blue) and doublet (red) crossings (circles). Increasing $U\equiv U_{\mathrm{L}}=U_{\mathrm{R}}$ shifts singlet crossings to $t_{\mathrm{LR}}\rightarrow 0$ until they vanish (a4). Doublet crossings show negligible $U$-dependence. 
        (b$1$)--(b$3$)
        Crossing values of $t_{\mathrm{LR}}$ at fixed $\varphi_{\mathrm{R}1}$ with varied $|\bm{\chi}|$ translate into singlet topological phase diagrams. 
        (c$1$)--(c$3$)
        Doublet phase diagrams have inaccessible regions (hatched) where crossings escape into the continuum.
		$(d1)$
		Triple dot with equal interdot hoppings $t$ supports the $\mathcal{P}_{2\pi/3}$ cyclic permutation symmetry. 
        (d$2$) 
        When ${\varphi_{\mathrm{M} 0}\!= \!4\pi/3}$ and ${\varphi_{\mathrm{M} 1}\!=\!\varphi_{\mathrm{R} 1} + 2\pi/3}$, the sBZ becomes 3D and $\mathcal{P}_{2\pi/3}$ can be satisfied on a 1D $\bm{\chi}$-manifold defined by Eq.~\eqref{eq:star}. 
		$(d3)$ 
        Sweeping $t$ tunes singlet (blue) and doublet (red) crossings (circles) on the $\bm{\chi}$-manifold ($|\bm{\chi}|=1/\sqrt{2}$). 
		(d$4$) 
        The resulting singlet phase diagram for the full interacting triple dot ($U/\Gamma=4.0$, $\delta/\Delta=0.5$) consists of two phases.
		\label{fig:dmrg}}
\end{figure*}

The doublet Hamiltonians~\eqref{eq:blocks} are identical for ${x=\uparrow,\downarrow}$. While off-diagonal blocks explicitly depend on $dU$, their specific algebraic structure permits analytic diagonalization even for $dU \neq 0$ when $\delta_{\mathrm L}=\delta_{\mathrm R}$. Each $S_z$ projection results in four eigenenergies 
$
\varepsilon_{\pm \pm}
\!=\!
2\delta \!-\! U/2
\pm
\left(
	\mu
	\!\pm\!
	\nu
\right)
$, 
where $\mu \!=\! \sqrt{\delta^2 \!+\! |\Gamma\bm{\chi}|^2} $ and $\nu \!=\!  \sqrt{|t_{\mathrm{LR}}|^2+dU^2}$. Crucially, sweeping $t_{\mathrm{LR}}$ on the $\bm{\chi}$-manifold induces isolated crossing between $\varepsilon_{+-}$ and $\varepsilon_{--}$ states at 
\begin{align}
	|t_{\mathrm{LR},\sigma}|
	=
	\sqrt{\delta^2+|\Gamma \bm{\chi}|^2 -dU^2}
	\label{eq:doub_cross}
\end{align}
as marked in Figs.~\ref{fig:al_results}(a1)-(a3) by red circles. Critically, while doublet crossings are independent of $U$ they disappear once $dU > \sqrt{\delta^2+|\Gamma \bm{\chi}|^2}$ as illustrated in Fig.~\ref{fig:al_results}(a4).

Eqs.~\eqref{eq:sing_cross} and~\eqref{eq:doub_cross} prescribe a unique $|\bm{\chi}|\!=\!|\bm{\chi}_{\mathrm L}|\!=\!|\bm{\chi}_{\mathrm R}|$ to a given $t_{\mathrm{LR}}$. The bands forming such isolated crossing on the $\bm{\chi}$-manifold are then projected onto an effective Weyl Hamiltonian $E_0\sigma_0+\sum_j d_j\sigma_j$, with $j \in \{x,y,z\}$ and $d_j$ parametrized by $(d\varphi_{x},d\varphi_{y},d\varphi_{z}) \equiv (d\varphi_{\mathrm{L} 1},d\varphi_{\mathrm{R} 0},d\varphi_{\mathrm{R} 1})$ measuring deviations from the crossing. The velocity matrices $V_{ab}\!=\!\partial d_a/\partial d\varphi_{b}$ return $\mathrm{det}(V)\!\neq\!0$ for $|\bm{\chi}|\!\neq\!0$ and $\delta\!\neq\!0$, thereby confirming linear gap opening in the entire sBZ while signs of $\mathrm{det}(V)$ assign topological charges $\pm1$ ($\pm2$) to singlet (doublet) crossings, the latter due to spin-rotation invariance. For $|\bm{\chi}|\!=\!0$, crossings occur either simultaneously (nodal lines) or not at all, so non-trivial topology arises only by varying $t_{\mathrm{LR}}$ at fixed $\varphi_{\mathrm{R} 1}$~\footnote{One defines a non-compact sBZ involving $\varphi_{\mathrm{L} 1}$, $\varphi_{\mathrm{R} 0}$, $t_{\mathrm{LR}}$ where velocity matrices are defined as $V_{ab}=\partial d_a/\partial du_{b}$ with $du_{x} \equiv d\varphi_{\mathrm{L} 1}$, $du_{y} \equiv d\varphi_{\mathrm{R} 0}$ and $du_{z} \equiv d t_{\mathrm{LR}}$.}. Small deviations from the case of all equal couplings remove these lines entirely (see SM).

Importantly, Eqs.~\eqref{eq:sing_cross} and~\eqref{eq:doub_cross} determine the topological phase diagrams in the $(\varphi_{\mathrm{R} 1}, t_{\mathrm{LR}})$ plane analytically, as shown in the End Matter. For equal dot-lead hybridizations [Figs.~\ref{fig:al_results}(b2)--(b4) for singlets, (c2) for doublets], the $|\bm{\chi}|\!=\!0$ nodal lines translate into constant-$t_{\mathrm{LR}}$ phase boundaries that simultaneously set the lower bound for topologically non-trivial phases. For singlets, the lower bound reads $|t_{\mathrm{LR},\mathrm{s}}|\!=\!\sqrt{\delta(\delta-U/2)}$ at $|\bm{\chi}|\!=\!0$ while the upper is $|t_{\mathrm{LR},\mathrm{s}}|\!=\!\sqrt{\delta^2+\Gamma^2-(U/2)\sqrt{\delta^2+\Gamma^2}}$ for $|\bm{\chi}|\!=\!1$. The former vanishes as $U\!\rightarrow\!2\delta$ [Fig.~\ref{fig:al_results}(b3)], the latter as $U\!\rightarrow\!2\sqrt{\delta^2+\Gamma^2}$, leaving the entire diagram trivial. Doublets are $U$-independent and lose topological phases only via the interaction asymmetry $dU$. Varying the hybridizations $\Gamma_{il}$ independently reshapes the $\bm{\chi}$-manifold and removes the $|\bm{\chi}|\!=\!0$ nodal lines, thereby qualitatively altering the phase diagrams (see SM).

\textit{Full interacting model.}---While both $\mathcal{P}_{\pi/2}$ and the resulting $\bm{\chi}$-manifold remain analytic at finite $\Delta$, the competition between Kondo screening and superconductivity makes crossing relations analogous to~\eqref{eq:sing_cross}~and~\eqref{eq:doub_cross} accessible only numerically~\cite{Moca-2021}. This requires a representation of frequency-dependent tunneling self-energies $\mathbb{\Sigma}_i$ via semi-infinite hopping chains that couple to their corresponding QDs. Among the many equivalent representations~\cite{Chin-2010}, we adopt that of Ref.~\cite{Liu-2016,MixingZenodo}, whose exponentially decaying hoppings permit controlled chain truncation and an efficient implementation of Density Matrix Renormalization Group (DMRG) \cite{PhysRevB.80.165117,PhysRevB.78.195317}.

We use $\delta \!\equiv\! \delta_{\mathrm L}\!=\!\delta_{\mathrm R}$, $\Gamma \!\equiv\! \Gamma_{\mathrm L} \!=\! \Gamma_{\mathrm{R}}$ and $U \!\equiv\! U_{\mathrm{L}} \!=\! U_{\mathrm{R}}$ on the $\bm{\chi}$-manifold to satisfy $\mathcal{P}_{\pi/2}$~\footnote{Crucially, in chain representation $\mathcal{P}_{\pi/2}$ becomes a generalized left-right mirror symmetry of the chain Hamiltonians, directly exposing the decomposition of the spin multiplets into subsectors of definite parity $P=\pm $.}. Sweeping $t_{\mathrm{LR}}$ using DMRG with adiabatic tracking (see SM) tunes singlet and doublet crossings [blue and red circles in Figs.~\ref{fig:dmrg}(a1)-(a4)]. Non-zero velocity matrix determinants (see SM) confirm these crossings as Weyl nodes of charge $\pm1$ for singlets and as doubly degenerate Weyl nodes with charge $\pm 2$ for doublets. In analogy to the AL, increasing $U$ removes singlet Weyl nodes [Fig.~\ref{fig:dmrg}(a4)]. A genuinely finite-$\Delta$ feature, by contrast, is the escape of the ABSs into the continuum as visible in Figs.~\ref{fig:dmrg}(a1)-(a4).

Each numerically obtained crossing value $t_{\mathrm{LR}}$ maps onto the analytic phase boundaries defined by Eq.~\eqref{eq:boundaries} in the End Matter. The AL boundaries become renormalized with increasing Coulomb interactions again removing non-trivial singlet phases [Figs.~\ref{fig:dmrg}(b1)-(b3)]. At finite $\Delta$, doublet crossings escape into the continuum near $|\bm{\chi}|\approx 1$, a process impossible in the AL but well-known in full models in both trivial~\cite{Zalom-2022} and non-trivial~scenarios \cite{Repin-2019esc}, making the corresponding phase regions inaccessible as denoted by hatched surfaces in Figs.~\ref{fig:dmrg}(c1)-(c3). 

Crucially, symmetries $\mathcal{P}_{\alpha}$ and their role in organizing topology extend to multi-dot cases, where they generalize into dot-permutations with embedded phase multiplication by $e^{i\alpha}$. For instance, the triple-dot array of Fig.~\ref{fig:dmrg}(d1) with all dots joined by a common hopping $t$ and six independent SC leads allows permutations of pairs of dot indices as well as cyclic permutations involving all three dots once $U \! \equiv \! U_{\mathrm{L}} \!=\! U_{\mathrm{R}} \!=\! U_{\mathrm{M}}$, ${\delta \!\equiv \! \delta_{\mathrm L} \!=\! \delta_{\mathrm R} \!=\! \delta_{\mathrm M}}$ and ${\Gamma \!\equiv\! \Gamma_{\mathrm L} \!=\! \Gamma_{\mathrm{R}} \!=\! \Gamma_{\mathrm M}}$. For example, pairwise transformation $\mathcal{P}_{\pi/2,\mathrm{M}}$ swapping $\mathrm{L} \leftrightarrow \mathrm{R}$ requires $\bm{\chi}_{\mathrm M}=0$ and operates as ${\mathcal{P}_{\pi/2,\mathrm{M}}^{\dagger} d^{\dagger}_{\sigma \mathrm{M} } \mathcal{P}_{\pi/2,\mathrm{M}} = e^{i\pi/2} d^{\dagger}_{\sigma \mathrm{M} }}$ on $i=\mathrm{M}$ and as~\eqref{eq:unitary} on $i \in \{\mathrm{L},\mathrm{R}\}$. Analogous pairwise transformations apply when $\bm{\chi}_{\mathrm L}=0$ or $\bm{\chi}_{\mathrm R}=0$. On the other hand, cyclic permutation $\mathcal{S}: \mathrm{L} \! \rightarrow \! \mathrm{R} \!\rightarrow \! \mathrm{M} \! \rightarrow \! \mathrm{L}$ requires a condition
\begin{align}
	\bm{\chi}
	\equiv
	\bm{\chi}_{\mathrm L}
	=
	e^{i 2\pi/3}\bm{\chi}_{\mathrm R}
	=
	e^{i 4\pi/3}\bm{\chi}_{\mathrm M}
	\label{eq:star}
\end{align}
as previously observed only empirically~\cite{Ohnmacht-2025r}. $\mathcal{P}_{2\pi/3}$ acts then as ${\mathcal{P}_{2\pi/3}^{\dagger} d^{\dagger}_{\sigma i} \mathcal{P}_{2\pi/3} = e^{i2\pi/3} d^{\dagger}_{\sigma \mathcal{S}(i)}}$.

Since symmetries ${\mathcal{P}_{\pi/2,i}}$ permute two dots they result in topological diagrams analogous to~Fig.~\ref{fig:dmrg}. To realize $\mathcal{P}_{2\pi/3}$ and its distinct phases, we set ${\varphi_{\mathrm{M} 0} \!=\! 4\pi/3}$ and ${\varphi_{\mathrm{M} 1} \!=\! \varphi_{\mathrm{R} 1} + 2\pi/3}$, so that $\varphi_{\mathrm{L} 1}$, $\varphi_{\mathrm{R} 0}$, $\varphi_{\mathrm{R} 1}$ span the sBZ. For the case of same dot-lead hybridizations, $\bm{\chi}$-manifold follows from Eq.~\eqref{eq:star} [Fig.~\ref{fig:dmrg}(d2)]. Sweeping $t$ at its $|\bm{\chi}|=1/\sqrt{2}$ point induces singlet and doublet crossings [blue and red circles in Fig.~\ref{fig:dmrg}(d3)]. Varying $\varphi_{\mathrm{R} 1}$ changes $|\bm{\chi}|$ and thus the crossing values of~$t$ that mark a single boundary in the topological phase diagram [Fig.~\ref{fig:dmrg}(d4)]. The parameters are set to display the onset of the removal of non-trivial topological phases by $U$.

\textit{Experimental detection.}---Double-dot systems have been realized experimentally in various settings, including Andreev molecules~\cite{Pillet-2019c, Matsuo-2023, Su-2017}, Cooper-pair splitters~\cite{Hofstetter-2009cps,Herrmann-2010}, and poor man's Majoranas~\cite{Zatelli-2024}. In such devices, the parameters controlling the symmetries $\mathcal{P}_{\alpha}$ are accessible to varying degrees. The SC phase differences are set in situ by flux with high accuracy~\cite{Pankratova-2020, Matsuo-2023, Antonelli2025h}. Gate control, supported by targeted fabrication, provides the tunneling symmetry $\Gamma_L=\Gamma_R$ required by $\mathcal{P}_{\pi/2}$~\cite{Su-2017}. Crucially, only the total hybridizations $\Gamma_i$ and not the individual dot-lead hybridizations must be controlled, while the gap closing at the $\bm{\chi}$-manifold itself facilitates such fine-tuning~\footnote{Similar fine-tuning is successfully performed in devices realizing poor man's Majoranas~\cite{Zatelli-2024}.}. Moreover, the interdot-tunneling sweeps of Figs.~\ref{fig:al_results}(a1)--(a4) and~\ref{fig:dmrg}(a1)--(a4) can be implemented as gate sweeps of the interdot coupling, giving direct control over the band crossings.

The interactions $U_i$, on the other hand, are controllable by the electrostatic confinement of the dot, but their asymmetry is largely predetermined during fabrication. This, however, is precisely the asymmetry against which the Weyl nodes generated by $\mathcal{P}_{\pi/2}$ are robust, as rooted in the AL approximation: the singlet Weyl nodes are strictly independent of $dU$ and the doublet nodes remain pinned to the $\bm{\chi}$-manifold even for $dU\!\neq\!0$. Finite-$\Delta$ corrections break this pinning, but only weakly. Even a sizable
asymmetry $\eta_U\!=\!|U_{\mathrm{L}}-U_{\mathrm{R}}|/(U_{\mathrm{L}}+U_{\mathrm{R}})\!=\!10\%$
opens a same-spin gap of only below $10^{-2}\Delta$~[see Figs.~\ref{fig:exp}(a1)--(a3) in the End Matter]. Moreover, in gate-defined 2D electron gas (2DEG) devices, electrostatic control of the dot confinement and of the tunnel barriers gives access to weakly interacting regimes~\cite{Wang-2023}. Interactions and their asymmetries might then be neglected leaving the symmetries $\mathcal{P}_{\alpha}$ fully flux- and gate-tunable in situ.

At finite $U$, the singlet nodes are lifted into the excited spectrum, as follows for instance from Eqs.~\eqref{eq:sing_cross} and~\eqref{eq:doub_cross}. In such cases, standard techniques of microwave ABS spectroscopy~\cite{Canadas-2022, Fatemi-2022, tenKate-2025} can still be applied to both the singlet and the doublet bands. Indeed, Figs.~\ref{fig:dmrg}(a1)--(a4) show that the parity mismatch with the ground state at the crossing can be countered by thermal excitations: the lowest singlet participating in the crossing has  $E_{\mathrm{ABS}}\approx\Delta/10$, which leaves it sufficiently populated for microwave
spectroscopy~\footnote{At $50\,$mK and $\Delta\approx200\,\mu$eV, about $1\%$ of the population resides in the excited singlet state, from which transitions to its crossing partner can be driven. A typical detection fingerprint can thus rely on a thermally assisted microwave signal. Another important mechanism is quasiparticle poisoning, which keeps the junction in the excited state.}. For the doublets, a similar scenario applies, with the crossing either matching the
ground-state parity [Fig.~\ref{fig:dmrg}(a2)] or relying on a low-lying doublet
excited state [Fig.~\ref{fig:dmrg}(a3)].

\textit{Conclusions.}---Emergent geometric symmetries $\mathcal{P}_{\alpha}$ represent the organizing principle of ABS band topology in QD-based Josephson devices. Through a two-stage mechanism, they confine crossings to analytically prescribed $\bm{\chi}$-manifolds of the sBZ while interdot coupling selects isolated topologically non-trivial crossing points on it. Symmetries $\mathcal{P}_{\alpha}$ become exclusively flux- and gate-controllable in non-interacting scenarios but extend towards fully interacting dots, where identical interactions are required. However, Weyl nodes generated by symmetries  $\mathcal{P}_{\alpha}$ show robust behavior against interaction asymmetries. Consequently, symmetries $\mathcal{P}_{\alpha}$ and the resulting $\bm{\chi}$-manifolds do not only explain the emergence of topology in QD-based systems but also translate into a general design principle for targeted fabrication and operation of gate-tunable MTJJ devices with non-trivial topology.

\textit{Acknowledgements.}---PZ and GD would like to thank Don Rolih, Kacper Wrze\'{s}niewski and Rok \v{Z}itko for helpful discussions. PZ acknowledges the support by the Czech Republic-Germany Mobility programme (ID:8J25DE001) and by the TERAFIT project - CZ.02.01.01/00/22\_008/0004594. Research of GD was co-funded by the European Union (Physics for Future – Grant Agreement No.~101081515). DCO and WB would like to thank Daniel Bobok, Tom\'a\v{s} Novotn\'y, Gleb Seleznev and Martin \v{Z}onda for useful discussions and acknowledge support by the DAAD PPP (Projekt-ID: 57753337) and the Deutsche Forschungsgemeinschaft (DFG; German Research Foundation) by SFB 1432 (Project No. 425217212).

\textit{Data availability.}---The data that support the findings of this article are openly available~\cite{2026-topo_zenodo}.

%


\clearpage
\appendix
\onecolumngrid
\begin{center}
	{\large\textbf{End Matter}}
\end{center}
\twocolumngrid

\subsection{Topological phase diagrams \label{sec:phase_diags} }

Same-spin ABS bands are generically gapped in the synthetic BZ, with Weyl nodes required at the boundaries dividing topologically distinct phases \cite{Murakami-2007, Murakami-2008, Riwar-2016, Meyer-2017b}. Locating these boundaries without prior knowledge of the singularities requires computationally demanding calculations like the Fukui-Hatsugai-Suzuki (FHS) method~\cite{Fukui-2005}. These, for example, precluded investigation of doubly degenerate Weyl cones in Ref.~\cite{Teshler-2023} already in the AL approximation. 

The symmetry $\mathcal{P}_{\pi/2}$ allows us to locate the boundaries fully analytically in AL while the DMRG calculations in the full model simplify to straightforward $t_{\mathrm{LR}}$ sweeps of Fig.~\ref{fig:dmrg}(a1)-(a4). The key is the analytic condition ${\bm{\chi}_{\mathrm L}=-\bm{\chi}_{\mathrm R}}$ which for the example of the same dot-lead couplings, i. e. $\Gamma_{\mathrm{L}0}\!=\!\Gamma_{\mathrm{L}1}\!=\! \Gamma_{\mathrm{R}0}\!=\!\Gamma_{\mathrm{R}1}\!=\!\Gamma$, prescribes $\varphi_{\mathrm{L} 1}$ and $\varphi_{\mathrm{R} 0}$ once $\varphi_{\mathrm{R} 1}$ is fixed. In such a case, three types of solutions can be  parametrized by $\varphi\in[0,2\pi]$ as
\begin{subequations}
	\begin{align}
		\varphi_{\mathrm{R} 1} &= \varphi,      &\, \, \,\, \varphi_{\mathrm{R} 0} &= \pi+\varphi,  &\, \, \,\, \varphi_{\mathrm{L} 1} &= \pi, \\
		\varphi_{\mathrm{R} 1} &= \varphi + \pi,     &\, \, \,\, \varphi_{\mathrm{R} 0} &= \pi,       &\, \, \,\, \varphi_{\mathrm{L} 1} &= \varphi, \\
		\varphi_{\mathrm{R} 1} &= \pi,          &\, \, \,\, \varphi_{\mathrm{R} 0} &= \pi \pm \varphi, &\, \, \,\, \varphi_{\mathrm{L} 1} &= \pm \varphi.
	\end{align}
    \label{eq:boundaries}
\end{subequations}

Each of the three solutions translates to a specific $|\bm{\chi}|$ which then conversely prescribes a crossing value of $t_{\mathrm{LR}}$ at given $\varphi_{\mathrm{R} 1}$. In the AL, analytic expressions~\eqref{eq:sing_cross} and~\eqref{eq:doub_cross} yield the required conversion for singlets and doublets respectively. In the full model, numerically obtained conversion table between $|\bm{\chi}|$ and $t_{\mathrm{LR}}$ is obtained from sweeping $t_{\mathrm{LR}}$ at varying $|\bm{\chi}|$ similar to Figs.~\ref{fig:dmrg}(a1)-(a4). 

We demonstrate the procedure in detail for singlets within the AL approximation. The first solution requires $|\bm{\chi}|\!=\!|\bm{\chi}_{\mathrm L}|\!=\!|\bm{\chi}_{\mathrm R}|\!=\!0$ since $\varphi_{\mathrm{L} 1}\!=\!\pi$ forces $\bm{\chi}_{\mathrm L}\!=\! 1+e^{i\pi}\!=\!0$. This translates via~\eqref{eq:sing_cross} into a boundary with constant 
\begin{align}
	t_{\mathrm{LR}}^{\mathrm{b1,s}}(\varphi_{\mathrm{R} 1})
	&=
	\sqrt{ \delta (\delta - U/2 )}.
\end{align}
The second solution has $\Gamma|\bm{\chi}|\!=\!\Gamma\sqrt{[1\!-\!\cos\varphi_{\mathrm{R} 1}]/2}
\!\equiv\! A_{\varphi_{\mathrm{R} 1}}$ and yields a bell-shaped boundary:
\begin{align}
	t_{\mathrm{LR}}^{\mathrm{b2,s}}(\varphi_{\mathrm{R} 1})
	&=
	\sqrt{
		\delta^2 
		+
		 A_{\varphi_{\mathrm{R} 1}}^2
		- 
		(U/2) \sqrt{(\delta^2+ A_{\varphi_{\mathrm{R} 1}}^2)}}.
\end{align}

\begin{figure}[t]
	\includegraphics[width=1.0\columnwidth]{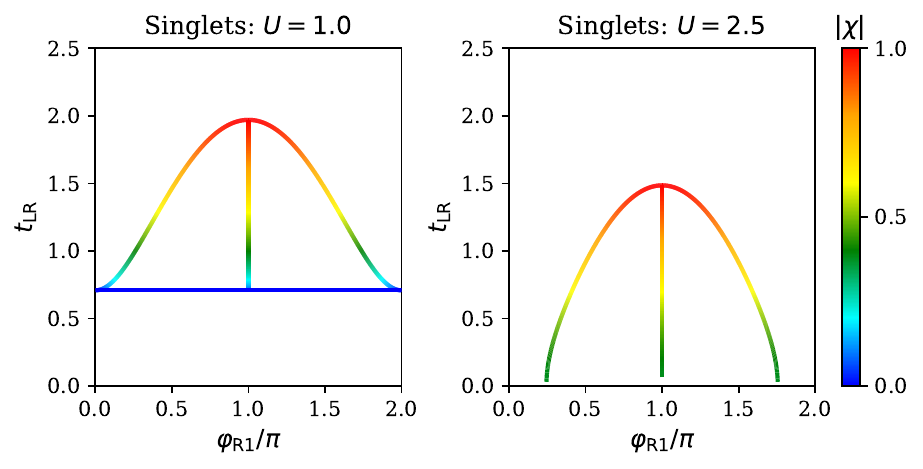}
	\vspace{-8mm}
	\caption{
		Topologically distinct phases in~model~\eqref{eq:Hdqd} require the emergence of boundaries where band gaps close at Weyl nodes. Using the $\bm{\chi}$-manifold defining feature ${\bm{\chi}_{\mathrm L}=-\bm{\chi}_{\mathrm R}}$ one can predict three distinct lines of Weyl nodes which obey Eq.~\ref{eq:boundaries}. In panels (a) and (b) we show these boundaries together with their $|\bm{\chi}|$-dependence according to the color bar.
		\label{fig:em}
	}
\end{figure}

The third solution encodes two different points in the sBZ at fixed $\varphi_{\mathrm{R} 1}\!=\!\pi$ but varying $\varphi_{\mathrm{L} 1}$. Consequently, $\Gamma|\bm{\chi}|\!=\!A_\varphi$ which then translates into
\begin{align}
	t_{\mathrm{LR}}^{\mathrm{b3,s}}(\varphi)
	&=
	\sqrt{
		\delta^2 
		+
		A_{\varphi}^2
		- 
		U \sqrt{(\delta^2+ A_{\varphi}^2)}/2}.
\end{align}

To calculate the corresponding Chern numbers, we count the overall number of Weyl nodes and their combined chirality (determined from the corresponding determinants of the velocity matrices in the sBZ) that appear on the phase boundaries~\cite{Herring-1937, Murakami-2007,Murakami-2008}. The second boundary yields for singlets only one Weyl node with positive chirality in the whole range of $\varphi_{\mathrm{R}1}$ which yields a jum of $+1$ in the Chern number. On the third boundary (vertical line), two Weyl nodes of the same negative chirality appear simultaneously. This results in a total jump of $-2$ for singlet topological phases. Finally, the first boundary has $|\bm{\chi}|=0$ which turns the band crossings into nodal lines in the sBZ spanned by $\varphi_{\mathrm{L}1}$, $\varphi_{\mathrm{R}0}$ and $\varphi_{\mathrm{R}1}$. However, to cross such horizontal boundaries in the the topological diagram one needs to, for example, fix $\varphi_{\mathrm{R}1}$ and vary $t_{LR}$. Moving across the boundary by varying $t_{LR}$ at fixed $\varphi_{\mathrm{R}1}$, however, requires us to characterize the band crossing with respect to the non-compact space spanned by  $\varphi_{\mathrm{L}1}$, $\varphi_{\mathrm{R}0}$ and $t_{LR}$. Here, the crossing turns out to be isolated with a positive (negative) chirality determined for $\varphi_{\mathrm{R}1}/\pi<1$ ($\varphi_{\mathrm{R}1}/\pi>1$). Finally, the two phases divided by the vertical line are required to have the same absolute value of the Chern number because of the underlying symmetry. Therefore, Chern numbers $-1,0,+1$ can be uniquely assigned to the observed singlet phases as shown in Figs.~\ref{fig:al_results}(b2)-(b4) and Figs.~\ref{fig:dmrg}(b1)-(b3). 

\begin{figure*}[t]
	\includegraphics[width=1.95\columnwidth]{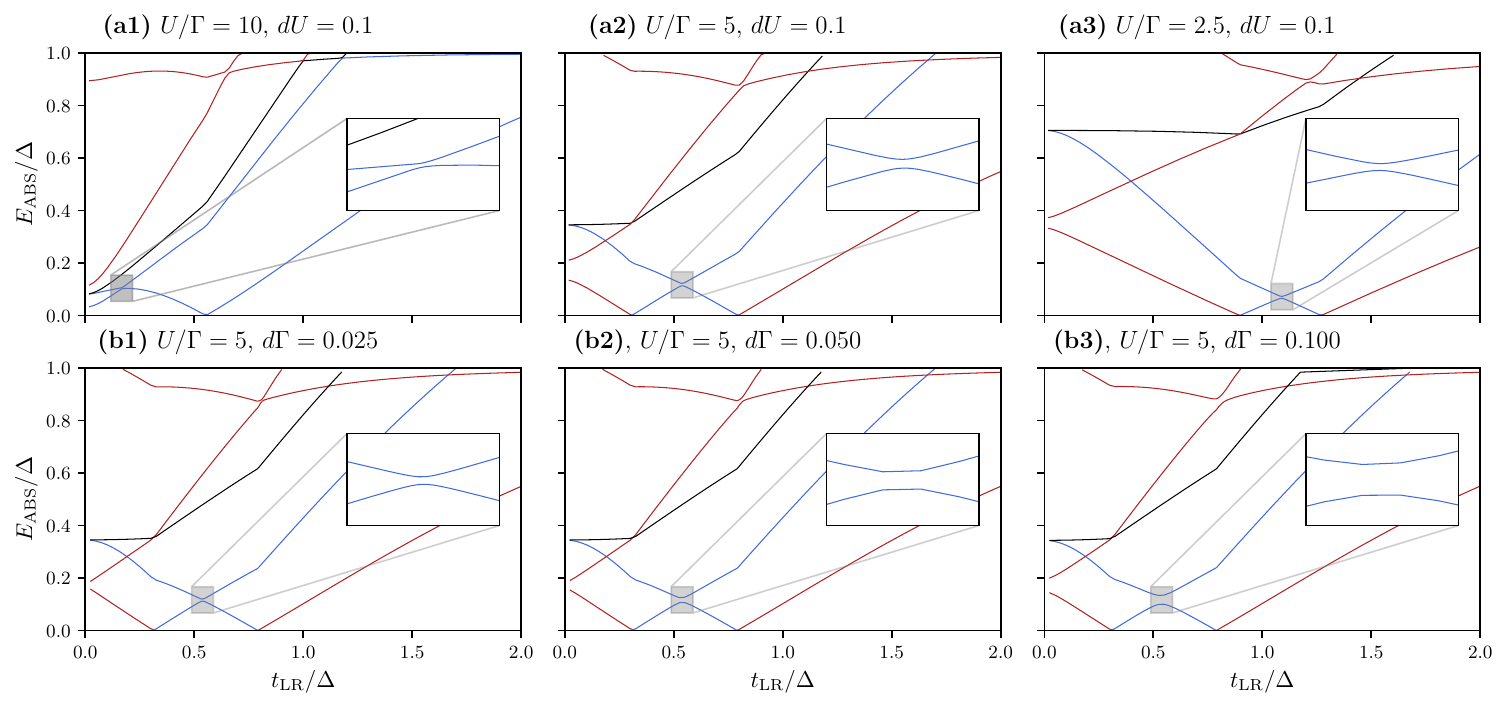}
	\vspace{-5mm}
\caption{
	Fate of the symmetry-pinned Weyl nodes when $\mathcal{P}_{\pi/2}$ is broken by asymmetries, obtained by adiabatic DMRG sweeps of $t_{\mathrm{LR}}$ at a fixed point of the $\bm{\chi}$-manifold ($\bm{\chi}_{\mathrm L}=-\bm{\chi}_{\mathrm R}$) with $U=2.0\,\Delta$ and $\delta_{\mathrm L}=\delta_{\mathrm R}$:
	(a1)--(a3)
	Interaction asymmetry $dU=0.1\,\Delta$ at increasing hybridization (i. e. $U/\Gamma=10$, $5$ and $2.5$) causes opening of the gap between same-spin ABS bands which is small compared to $\Delta$ and is controlled by $\Gamma$.
	(b1)--(b3)
	Tunneling asymmetry $\eta_{\Gamma}=|\Gamma_{\mathrm L}-\Gamma_{\mathrm{R}}|/(\Gamma_{\mathrm L}+\Gamma_{\mathrm{R}})=0.025$, $0.050$ and $0.100$ at $dU=0$ and $U/\Gamma=5$, opens the band gap monotonically and considerably faster than $dU$.
	\label{fig:exp}
}
\end{figure*}

For the doublet sector, analogous calculations allow us to translate the three solutions into the phase boundaries  via~\eqref{eq:doub_cross}. The resulting expressions read
\begin{align}
	t_{\mathrm{LR}}^{\mathrm{b1,d}}(\varphi_{\mathrm{R} 1})
	&=
	\delta,
	\\
	t_{\mathrm{LR}}^{\mathrm{b2,d}}(\varphi_{\mathrm{R} 1})
	&=
	\sqrt{
		\delta^2 
		+
		A_{\varphi_{\mathrm{R} 1}}^2		
	},
    \\
    t_{\mathrm{LR}}^{\mathrm{b3,d}}(\varphi)
    &=
    \sqrt{
    	\delta^2 
    	+
    	A_{\varphi}^2
    },
\end{align}
where $A_\varphi \!\equiv\! \Gamma \sqrt{[1-\cos\varphi]/2}$. The structure of boundaries mirrors the singlet case, with the first being a constant $t_{\mathrm{LR}}$ line, the second having a bell shape, and the third forming a vertical line at $\varphi_{\mathrm{R} 1}\!=\!\pi$ that separates two topologically non-trivial phases. However, each doublet-doublet crossing is doubly degenerate via spin rotation invariance and it carries the total charge $\pm2$ , assigning thus Chern numbers $-2,0,+2$ to the enclosed doublet phases as shown in Fig.~\ref{fig:al_results}(c2) and Figs.~\ref{fig:dmrg}(c1)-(c3).

\subsection{Effect of asymmetries in realistic applications\label{app:asymmetries}}

The symmetry $\mathcal{P}_{\pi/2}$ requires $\bm{\chi}_{\mathrm L}=-\bm{\chi}_{\mathrm R}$, $\delta_{\mathrm L}=\delta_{\mathrm R}$, $\Gamma_{\mathrm L}=\Gamma_{\mathrm{R}}$ and $U_{\mathrm{L}}=U_{\mathrm{R}}$, conditions posing challenges of differing severity in real devices. The phase differences $\varphi_{il}$ entering $\bm{\chi}_{\mathrm L}=-\bm{\chi}_{\mathrm R}$ are flux-controlled in situ with high accuracy, while $\Gamma_i$ and $\delta_i$ are gate-tunable but cross-correlated. The control of interactions and their asymmetries is instead platform dependent. In self-assembled and nanowire dots $U$ is sizable, but double dots with matched $U_{\mathrm{L}}\approx U_{\mathrm{R}}\sim1$--$2$\,meV$\,>\Delta$ have already been realized in QD--SC--QD geometries~\cite{Su-2017}. In gate-defined 2DEG devices $U$ is set by depletion with weakly confined dots reaching $U\ll\Delta$. Consequently, selecting a proper regime in a 2DEG platform the requirement on interaction symmetry can be completely sidestepped, making symmetry $\mathcal{P}_{\pi/2}$ exclusively flux- and gate-controllable. 

We now quantify the impact of asymmetries on the pinning of Weyl nodes to the $\bm{\chi}$-manifold. Starting with interaction asymmetries, we note that AL predict singlets as independent of $dU$. In the doublet sectors $dU$ does appear in the off-diagonal blocks of~\eqref{eq:blocks} and breaks $\mathcal{P}_{\pi/2}$ explicitly. However, the doublet AL Hamiltonian possesses a special algebraic structure permitting diagonalization in a basis of definite parity $P$. The resulting crossings~\eqref{eq:doub_cross} together with their velocity matrices (see SM) confirm that the $\bm{\chi}$-manifold still pins the doublet Weyl nodes despite  $\mathcal{P}_{\pi/2}$ broken by $dU\neq0$. Consequently, AL protects pinning of Weyl nodes effectively against interaction asymmetry.

Although the AL is an approximation, its predictions are known to reproduce realistic spectra once effective parameters are used~\cite{Grove-Rasmussen-2018}, which imposes stringent qualitative constraints on the full model~\eqref{eq:Hdqd}. Consequently, strong suppression of the gap between same-spin ABS bands is expected at the $\bm{\chi}$-manifold once interaction asymmetries are present at finite $\Delta$. To quantify this, we impose a sizable asymmetry $dU=0.1\,\Delta$ at $U=2.0\,\Delta$. This corresponds to a relative mismatch of $\eta_U\!=\!|U_{\mathrm{L}}-U_{\mathrm{R}}|/(U_{\mathrm{L}}+U_{\mathrm{R}})\!=\!10\%$ which represents a highly conservative upper bound on fabrication imperfections. Sweeping $t_{\mathrm{LR}}$ with adiabatic DMRG at $\bm{\chi}$-manifold [Fig.~\ref{fig:exp}$(a1)$--$(a3)$], the same-spin gap reaches only $10^{-2}\Delta$ in the least favorable case ($U/\Gamma=5$) and drops to $4\times10^{-3}\Delta$ at $U/\Gamma=10$. Tunneling asymmetry $\eta_{\Gamma}=|\Gamma_{\mathrm L}-\Gamma_{\mathrm{R}}|/(\Gamma_{\mathrm L}+\Gamma_{\mathrm{R}})$ opens the gap monotonically and considerably faster, more slowly for doublets than for singlets [Fig.~\ref{fig:exp}$(b1)$--$(b3)$]. This again traces back to the AL, where $\eta_{\Gamma}\neq0$ displaces the Weyl nodes to a redefined $\bm{\chi}$-manifold obeying $\Gamma_{\mathrm L}\bm{\chi}_{\mathrm L}=-\Gamma_{\mathrm{R}}\bm{\chi}_{\mathrm R}$ (see SM).

Taken together, asymmetries in the full model break $\mathcal{P}_{\pi/2}$ in every spin sector -- including the singlets, which the AL leaves untouched -- so the $\bm{\chi}$-manifold ceases to pin the Weyl nodes exactly. Consequently, a gap between same spin ABS bands opens at the $\bm{\chi}$-manifold once asymmetries are present. However, such gaps remain small and gate-tunable when imposed by interaction asymmetries as anticipated by the AL.

\end{document}